\documentclass[11pt]{article}
\usepackage[utf8]{inputenc}
\usepackage{geometry}
\usepackage{amssymb}
\usepackage{upgreek}
\usepackage{setspace}
\usepackage{amsmath}
\usepackage{makecell}
\usepackage{multirow}
\usepackage{adjustbox}
\usepackage{url}
\usepackage{booktabs}
\usepackage{algorithm}
\usepackage{algpseudocode}
\newcommand{\mb}{\mathbf}
\newcommand{\bs}{\boldsymbol}

\usepackage{verbatim}
\usepackage{caption}

\usepackage[compact]{titlesec}
\titlespacing{\section}{0pt}{2ex}{1ex}
\titlespacing{\subsection}{0pt}{1ex}{0ex}
\titlespacing{\subsubsection}{0pt}{0.5ex}{0ex}

\usepackage{natbib}
\usepackage{graphicx}

\usepackage{xr}
\makeatletter
\newcommand*{\addFileDependency}[1]{
  \typeout{(#1)}
  \@addtofilelist{#1}
  \IfFileExists{#1}{}{\typeout{No file #1.}}
}
\makeatother

\begin{document}

\thispagestyle{empty} \baselineskip=28pt \vskip 5mm
\begin{center}
{\Huge{\bf  CESAR: A Convolutional Echo State AutoencodeR for High-Resolution Wind Forecasting}}
\end{center}

\baselineskip=12pt \vskip 10mm

\begin{center}\large
Matthew Bonas\footnote[1]{
\baselineskip=11pt Department of Applied and Computational Mathematics and Statistics,
University of Notre Dame,
Notre Dame, IN 46556, USA.},
Paolo Giani\footnote[2]{Department of Earth, Atmospheric and Planetary Sciences, 
Massachusetts Institute of Technology,
Cambridge, MA 02139, USA.},
Paola Crippa\footnote[3]{Department of Civil and Environmental Engineering and Earth Sciences, University of Notre Dame,
Notre Dame, IN 46556, USA.} and
Stefano Castruccio\textsuperscript{1}
\end{center}

\baselineskip=17pt \vskip 10mm \vskip 10mm

\begin{center}
{\large{\bf Abstract}}
\end{center}

\doublespacing

An accurate and timely assessment of wind speed and energy output allows an efficient planning and management of this resource on the power grid. Wind energy, especially at high resolution, calls for the development of nonlinear statistical models able to capture complex dependencies in space and time. This work introduces a Convolutional Echo State AutoencodeR (CESAR), a spatio-temporal, neural network-based model which first extracts the spatial features with a deep convolutional autoencoder, and then models their dynamics with an echo state network. We also propose a two-step approach to also allow for computationally affordable inference, while also performing uncertainty quantification. We focus on a high-resolution simulation in Riyadh (Saudi Arabia), an area where wind farm planning is currently ongoing, and show how CESAR is able to provide improved forecasting of wind speed and power for proposed building sites by up to 17\% against the best alternative methods. 

\newpage

\section{Introduction}\label{sec:intro}

Since 1950, global consumption of fossil fuels has increased eightfold, from 2,500 million tonnes of oil equivalent (Mtoe) per year to 20,000 million tonnes of Mtoe \citep{ren24}. As oil is a finite resource, its rapid depletion poses a significant challenge for the future, underscoring the need for alternative, clean, renewable energy sources. Wind energy, the focus of this study, has seen substantial growth in recent decades and has already contributed to reductions in greenhouse gas emissions and local air pollution. As of 2024, wind energy contributes approximately 6-7\% of global electricity generation, continuing its growth trajectory from 5\% in 2019, with China and the United States being the two largest producers \citep{gwec24}. Despite the global growth in wind energy, regions such as the Middle East and North Africa still generate very little wind power \citep{moh21}. Saudi Arabia, for example, remains almost entirely dependent on its vast oil reserves for domestic energy needs but has recently begun developing plans to diversify its energy mix with renewables. As part of its Vision 2030 initiative, the country aims to generate 16 GW of wind power, positioning it as a leader in wind energy production \citep{nur17}. Given the strategic importance of wind energy in the present and future of Saudi Arabia, and more broadly globally, a key priority is to develop methods to accurately predict it to allow optimal allocation in the energy grid.

Traditional statistical techniques for modeling wind cater from the time series literature, which in their simplest form predicate linear dependence from past observations: models such as the autoregressive integrated moving average (ARIMA, \cite{brockwell16}) have been widely used due to their simplicity and effectiveness \citep{zhang03}. However, the complex, nonlinear nature of wind speed (especially at high resolutions) calls for the development of more flexible models. The current practice to model temporal and spatio-temporal data, as demonstrated by the decades of literature on the topic \citep{wik19}, is to formulate them in a hierarchical framework. Under this paradigm, the data are independent conditional to a latent spatio-temporal process, whose dynamics is then expressed through a time series model. When turbines are not operational but only planned, a forecast over a possibly large candidate area is necessary, thereby motivating the adoption of methods which are not just flexible, but also scalable. In this regard, the aforementioned hierarchical framework assumes that the latent process has a reduced spatial representation, and a broad large of models have been developed depending on the type of spatial reduction performed, from fixed rank kriging (FRK, \cite{cre08}) to finite elements-based methods based on stochastic partial differential equations \citep{lin11}. All these class of models have been very popular, yet the functional form of the spatial reduction has been almost exclusively limited to a (stochastic) linear function, with appropriately chosen knots and dependence structure over a reduced space.

The machine learning literature has also independently explored both the topic of spatial feature extraction for large data and dynamic modeling, especially with constructs leveraging on neural networks. For spatial feature extraction, \textit{convolutional autoencoders} (CAE, \cite{kramer91, kramer92}) is a popular class of methods based on a convolutional neural network learning an efficient, compressed representation of the data. CAEs consist of an encoder that maps the data to a latent space and a decoder that reconstructs it from this compressed representation, thereby allowing to effectively reduce the dimensions of the input data \citep{wang14, wang16}. For modeling data in time, recurrent neural network models and generalizations such as long short-term memory networks \citep{liu14, xue23} are the current standard. Instabilities when performing inference due to gradient computation, along with the challenges in identifiability for relatively short time series has prompted the development of alternative approaches predicated on stochastic weight matrices. These \textit{echo state networks} (ESNs, \cite{jae01, jae07}) have been shown to be more flexible in capturing nonlinear dynamics \citep{bona24}. In particular their stochastic nature makes their formulation a statistical model very natural \citep{mcd17, mcd19a, mcd19b,yoo23} and allowed the development of calibration approaches \citep{bonas23}, ensemble models \citep{bonas2024a}, non-Gaussian models \citep{moncada24} and generalization to graphical neural networks \citep{wan24}. In the case of spatio-temporal forecasting for large datasets, ESNs have been traditionally coupled with kriging-based dimension reduction approaches \citep{huang2021}. Spatial and spatio-temporal methods in the machine learning community have therefore large potential due to the additional flexibility offered by neural networks, yet their formulation so far has been predominantly algorithmic and fails to recognize the existence of a model-based approach which can, crucially, provide formal uncertainty quantification. 

In this work, we propose a \textit{Convolutional Echo State AutoencodeR} (CESAR), a neural network-based approach for both spatial feature extraction with a CAE and dynamic modeling with an ESN which is able to leverage the dimension reduction and feature extraction abilities of a CAE while also producing accurate forecast estimates using an echo state network architecture. Crucially, we frame CESAR as a nonlinear generalization of a hierarchical statistical model, thereby bridging the gap between machine learning models and spatio-temporal statistics models. We will also propose a computationally efficient inference approach which, by virtue of the model-based framework that we formulate, will allow to perform (calibrated) uncertainty quantification using ensemble-based methods. 

Saudi Arabia does not have a structured industry for operational wind power forecasting relying on weather models such as the High-Resolution Rapid Refresh model \citep{dow22} as in the United States. As is often the case in emerging economies, it is instead necessary to rely on ground observational data, publicly available coarse data, or perform a proof-of-concept high-resolution simulations. Early studies on the country's wind energy resources have relied on available data from the Middle East North Africa Coordinated Regional Climate Downscaling Experiment (MENA CORDEX) experiment \citep{che18} or the NCAR large ensemble \citep{tag19,jeo19}. More recently, a targeted high-resolution study across the country with the Weather and Research Forecasting (WRF) was performed \citep{giani20}, allowing a much more precise assessment of the current and future resources \citep{tag20,che21,cri21,zha21,huang2021}, and in particular highlighting the region near Riyadh as a candidate for wind turbine construction. In this work, we will use a recent simulation performed at very high-resolution ($<$1km) in this region \citep{giani2022} to understand the accuracy of CESAR in predicting energy output from planned turbines. The proposed method represents the first template for an operational forecasting in the Riyadh region once the planned turbines will be active on the energy grid.

The manuscript proceeds as follows. In Section \ref{sec:data} we introduce the wind data over Riyadh, Saudi Arabia. In Section \ref{sec:methods} we describe the methodology. In Section \ref{sec:simstudy} we present a simulation study to assess the ability of our proposed approach to produce forecasts while also properly quantifying the uncertainty. Section \ref{sec:appl} applies the proposed approach to the wind speed data and details a method to estimate the future wind power from the forecasts. Section \ref{sec:discuss} concludes with a discussion.

\section{Data Description}\label{sec:data}

We use high-resolution 10m wind speed data simulated with the Weather Research and Forecasting (WRF) model over the Riyadh region in Saudi Arabia. WRF is a numerical weather prediction system that solves the nonhydrostatic compressible Euler Equations to calculate atmospheric properties such as air temperature, pressure, humidity and wind speed \citep{skamarock2008}. The dataset was generated by performing coupled meso- to micro-scale simulations with WRF, an approach that allows to reproduce microscale flow properties with high fidelity through a dynamical downscaling approach \citep{giani2022, giani2024} and that can be used to draw conclusions on wind power generation \citep{demoliner2024}. 
Details on the model setup and physics options adopted can be found in \cite{giani2022} (simulation labeled REF), here we summarize some of the key features. 

The model includes high-resolution surface properties (e.g., terrain elevation), as well as large-scale atmospheric forcing from the European Centre for Medium-Range Weather Forecast (HRES-ECMWF) reanalysis data \citep{european2011ecmwf}. Initial and boundary conditions are set by the parent lower resolution simulations via the nesting procedure, where the outermost domain is forced by HRES-ECMWF. The model solves the dynamical equations, along with parameterizations of radiation, surface fluxes of heat, moisture and momentum, turbulent transfer, clouds and precipitation, to compute the space-time fields of several atmospheric properties, including the surface zonal ($u_{10}$) and meridional ($v_{10}$) components of wind that we use here, where the subscript denotes 10 meters above ground level in this context. For this work we consider the wind speed (magnitude) $\mb{X} = \sqrt{u_{10}^2 + v_{10}^2}$, measured in meters per second ($ms^{-1}$).

The simulation covers 10 dry summer days (July 22, 2016 to August 1, 2016) in the Riyadh region with high spatial resolution (horizontal grid spacing $\Delta x \approx 450$ meters). The grid cells are distributed across a $m\times n$ grid, with $m=n=256$ for a total of $65,536$ spatial locations over an area of approximately 115$\times$115 km$^2$. We consider hourly wind speed from the atmospheric simulation, which results in $T=24\times10=240$ time points for each location. The total dataset size is therefore $m\times n\times T \approx 15.7$ million data points. 

\begin{figure}[!bt]
\centering
\includegraphics[width = 14cm]{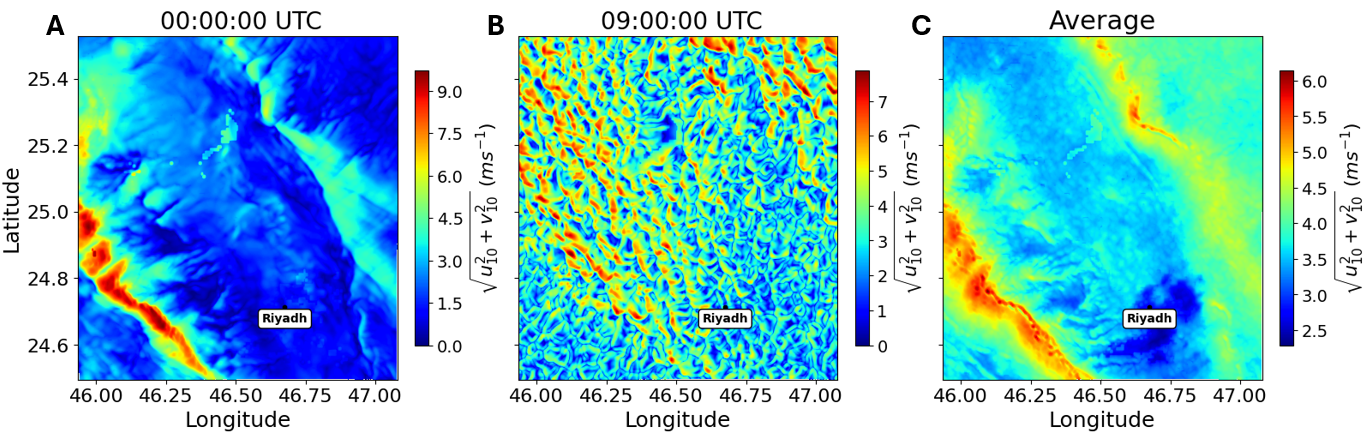}
\caption{Wind speed (ms$^{-1}$) output from the WRF model over Riyadh, Saudi Arabia for (A) nighttime and (B) daytime on 2016-07-26. Panel (C) shows the average hourly wind speed across all 10 days from 2016-07-22 through 2016-08-01}
\label{fig:WSDataPlot}
\end{figure}

Figure \ref{fig:WSDataPlot} shows how the wind field structure changes considerably between the turbulent daytime and the stable nighttime. During the afternoon, the atmosphere is in a turbulent state because of convection that transports surface heat up to the top of the atmospheric boundary layer. Figure \ref{fig:WSDataPlot}B shows the convective turbulence structure (at noon local time of July 26, 2016), which can be observed due to the very fine spatial resolution of the simulation ($<$1km). Conversely, Figure \ref{fig:WSDataPlot}A shows the stable boundary layer that forms at night, where no convective turbulence is present due to the lack of solar radiation. The time-averaged wind speed (Figure \ref{fig:WSDataPlot}C) is follows the local topography (Figure 1 in \citep{giani2022}), with higher winds generally corresponding to higher terrain elevation.

\section{Methodology}\label{sec:methods}

Throughout this Section we introduce the model and the approaches for inference and uncertainty quantification. Specifically, we introduce the general spatio-temporal framework in Section \ref{sec:spacetime}. In Section \ref{sec:spatialrecon} we explain the spatial part of the model, i.e., how the CAE extracts the data features. In Section \ref{sec:forecasting} we describe the temporal model applied to the extracted CAE features. In Section \ref{sec:Inference} we detail the inference and finally in Section \ref{sec:calibration} we detail the approach to uncertainty quantification.

\subsection{Hierarchical Spatio-Temporal Models}\label{sec:spacetime}

For simplicity we assume Gaussian data, even though the proposed framework can easily be extended to any non-Gaussian distribution from the exponential family. Also, for stability we scale the data via min-max normalization (but we denote them with the same notation for convenience), and we operate under the assumption that there are no covariates. 

The ultimate goal of this work is to forecast high-resolution spatio-temporal data given their past history. Formally, we consider $p$ variables of interest observed on a $m \times n$ grid and organized as $\mb{X}_{t} \in \mathbb{R}^{m \times n \times p}$ at times $t=1,\dots,T$. We aim to forecast $\mb{X}_{T+\tau}$ for some integer $\tau>0$, as well as its distributions given $\mb{X}_{T},\mb{X}_{T-1}, \dots, \mb{X}_{1}$. From a high level modeling perspective, we follow the conventional hierarchical framework of \cite{wikle06} which specifies the data conditional on a latent spatio-temporal process, and then decouples the spatial and temporal dependence according to a state-space model \citep{durbin12}. Formally, we assume the following:
\begin{subequations}\label{eqn:statespace}
    \begin{flalign}
        \text{Observation Equation:} & \qquad \mb{X}_{t} = h_o\left(\mb{Y}_{t} \mid \bs{\theta}_o\right) + \bs{\eta}_{t}, \quad \bs{\eta}_{t} \thicksim \mathcal{N}(\mb{0}, \sigma^2_{o}\bs{I}),\label{eqn:space} \\
        \text{State Equation:} & \qquad \mb{Y}_{t} = h_s\left(\mb{Y}_{t-1}, \ldots, \mb{Y}_{1}\mid \bs{\theta}_{s}\right) + \bs{\upsilon}_{t}, \quad \bs{\upsilon}_{t} \thicksim \mathcal{N}(\mb{0}, \sigma^2_{s}\bs{I}),\label{eqn:time}
    \end{flalign}
\end{subequations}

\noindent where $\mb{Y}_{t}$ represents the latent state at time $t$. Under this framework $\bs{\eta}_{t}$ represents the measurement noise and $\bs{\upsilon}_{t}$ the process noise, both of which are zero mean multivariate Gaussian distributed with diagonal covariance matrices and variances $\sigma^2_{o}$ and $\sigma^2_{s}$. Both $h_o(\cdot)$ and $h_s(\cdot)$ represent (possibly stochastic) functions dependent on some parameters $\bs{\theta}_{o}$ and $\bs{\theta}_{s}$, respectively. 

A convenient modeling strategy to explain the relationship between the observed data and the latent space is to assume that $h_o$ is a linear function. This approach drives common feature extraction approaches such as principal component analysis (PCA, \cite{joll14}) or empirical orthogonal functions (EOFs, \cite{hann07}). More spatially-aware approaches predicate reducing the dimension by selecting a set of fixed representative locations (or knots) and reconstruct the original data linearly and conditionally from these. Once the spatial features have been extracted, they can then be modeled with some time series model such as ARIMA or generalizations thereof. 

In this work, we assume that both $h_o$ and $h_s$ in \eqref{eqn:statespace} are nonlinear and controlled by deep neural networks, a solution which allows flexibility in capturing high resolution wind speed and power.

\subsection{Space: Convolutional Autoencoders}\label{sec:spatialrecon}

Since in this Section we detail the spatial model, we omit the temporal subscript for notational simplicity, and we index the data as $\mb{X}=\{X_{i,j,k}, i=1, \ldots, m, j=1, \ldots, n; k=1, \ldots, p\}$, where $(i,j)$ represents the spatial location on a grid and $k$ represents the variable of interest. CAEs are deep learning models designed for unsupervised learning tasks in image processing which are able to learn representations of high-dimensional data by leveraging on their spatial structure \citep{vincent10, masci11}. 

\begin{figure}[!tb]
\centering
\includegraphics[width = 12.5cm]{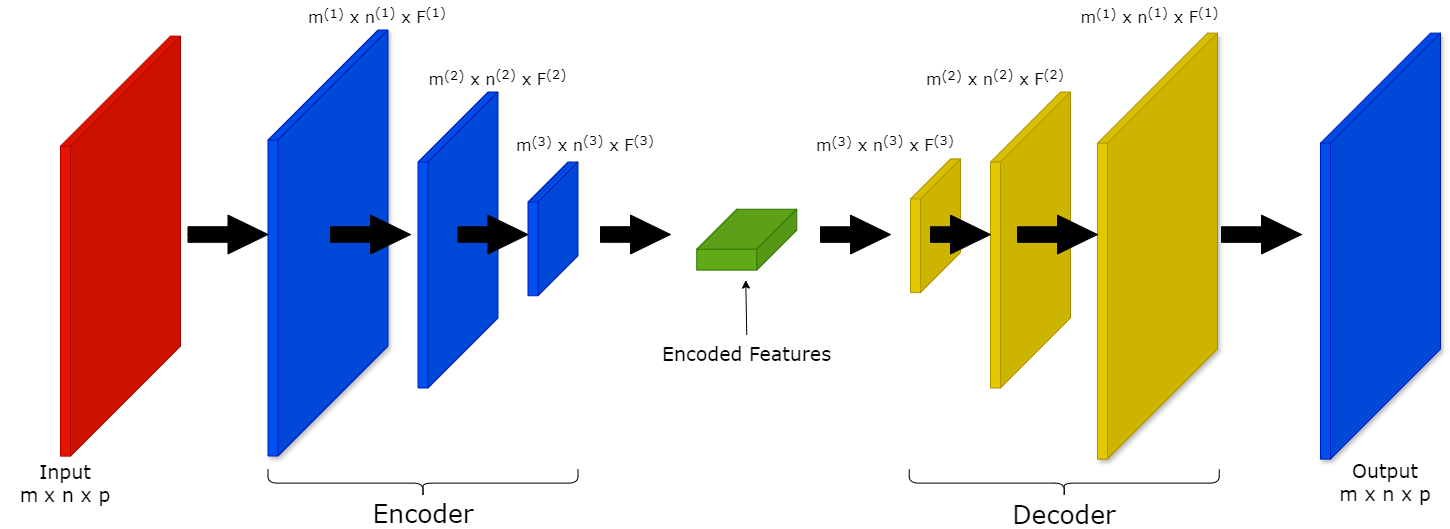}
\caption{Schematic for a convolutional autoencoder for two dimensional. Red represents the input data, blue represents convolutional layers, yellow represents deconvolutional layers and green here represents the encoded features/feature maps.}
\label{fig:AEArchitecture}
\end{figure}

The key principle of a CAE is to have a function $h_o$ in \eqref{eqn:space} which is a deep convolutional neural network to extracts the spatial features. Differently from canonical dimension reduction techniques such as PCA or FRK, the mapping from the data $\mb{X}$ to the latent space $\mb{Y}$ is not the same as the (inverse) mapping $\mb{Y}$ to $\mb{X}$. Specifically, a CAE consists of an \textit{encoder} and a \textit{decoder}, where the encoder compresses the input $\mb{X}$ into a latent representation $\mb{Y}$ using $L$ convolutional layers, each of which reduces the data size (depicted in blue in Figure \ref{fig:AEArchitecture}). For each layer $\ell = 1,\dots,L$ one specifies a number $\mathcal{F}^{(\ell)}$ of $k \times k$ (with $k \ll n,m$) matrices (\textit{filters}) with unknown parameters $w^{(\ell)}_{a,b;\tilde{f},f}$, as well as a bias terms $w^{(\ell)}_{0;f}$. Among the four filter subscripts, $a,b=1, \ldots k$ denote the row and columns entries, $\tilde{f}=1, \ldots, \mathcal{F}^{(\ell-1)}$ refers to a filter in the previous layer $\ell-1$ and $f=1, \ldots, \mathcal{F}^{(\ell)}$ refers to the filter in the current layer $\ell$. The filters are used to perform a local convolution of the input every $\xi$ elements in both horizontal and vertical directions (\textit{strides}), and then the result is summed across all the filters from the previous layer. This allows to capture local patterns such as edges and textures, and greatly reduces the input dimensionality since the convolution requires only $k^2$ parameters per filter, a small number compared to the original $n\times m$ data size. 

Formally, for a layer $\ell$, the convolution results in a feature map $\mathbb{R}^{m^{(\ell)} \times n^{(\ell)} \times \mathcal{F}^{(\ell)}} \ni \mb{Y}^{(\ell)}=\{Y^{(\ell)}_{i,j;f}, i = 1,\dots,m^{(\ell)}, j= 1,\dots,n^{(\ell)}, f = 1,\dots,\mathcal{F}^{(\ell)}\}$ tensor, where $m^{(\ell)} \times n^{(\ell)}$ denotes the size of the $\mathcal{F}^{(\ell)}$ output feature map, and is defined as:
\begin{equation}\label{eqn:encoder}
    Y^{(\ell)}_{i,j;f} = g\left(\sum_{\tilde{f}=1}^{\mathcal{F}^{(\ell-1)}} \sum_{a, b=1}^k Y^{(\ell-1)}_{\xi i+a,\xi j+b;\tilde{f}} \cdot w^{(\ell)}_{a,b;\tilde{f},f} + w^{(\ell)}_{0;f} \right).
\end{equation}

\noindent In the first layer we consider the original data, so for $\ell = 1$ we have that $\mathcal{F}^{(0)} = p$ representing the number of variables in the data, and $Y^{(0)}_{i,j;\tilde{f}} = X_{i,j,p}$. When $\ell = L$ (the last layer), equation \eqref{eqn:encoder} provides the extracted feature vector $\mb{Y}^{(L)}=\mb{Y}$ (green block in Figure \ref{fig:AEArchitecture}). The function $g(\cdot)$ is some nonlinear function (\textit{activation function}) such as rectified linear \citep{goo16}.  

The second component of the CAE is the decoder block (in yellow in Figure \ref{fig:AEArchitecture}), which reconstructs the original data $\mb{X}$ from the encoded feature $\mb{Y}^{(L)}$. The decoder consists of an equivalent number of $L$ deconvolutional layers using the same stride $\xi$ as the encoder. The number of filters per layer $\mb{\mathcal{F}}^{\prime}$ in the decoder is the same as in the encoder, except the order is reversed. For example, if the encoder block has $L=3$ convolutional layers with filters $\mathcal{F} = \{4,8,16\}$ then the number of filters in the corresponding $L=3$ deconvolutional layers is $\mathcal{F}^{\prime} = \{16,8,4\}$. These layers reverse the operations performed by the encoder, restoring the dimensions and the input data. Formally, the feature maps $\mb{Y}^{\prime (\ell)}=\{Y^{\prime (\ell)}_{i,j;f^{\prime}}, i = 1,\dots,m^{(\ell)}, i = 1,\dots,n^{(\ell)}, f^{\prime} = 1,\dots,\mathcal{F}^{\prime (\ell)}\}$ from each decoder layer $\ell$ is obtained the previous layer $\mb{Y}^{\prime (\ell-1)}$ as:
\begin{equation}\label{eqn:decoder}
    Y^{\prime (\ell)}_{i,j;f^{\prime}} = g\left(\sum_{\tilde{f}=1}^{\mathcal{F}^{\prime (\ell-1)}} \sum_{a, b=1}^k Y^{^\prime(\ell-1)}_{\lfloor\frac{i}{\xi}\rfloor+a,\lfloor\frac{j}{\xi}\rfloor+b;\tilde{f}} \cdot w^{\prime (\ell)}_{a,b;\tilde{f},f^{\prime}} +  w'^{(\ell)}_{0;f} \right),
\end{equation}

\noindent where $\lfloor \cdot \rfloor$ is the floor operator and $g(\cdot)$ is the activation function. As in the encoder case, $w^{\prime}_{a,b;\tilde{f},f^{\prime}}$ denotes the parameters for some filter $f^{\prime (\ell)}$, while $w'^{(\ell)}_{0;f}$ represents the bias.  In the case where $\ell = 1$ then $Y^{\prime (1)}_{i,j;f^{\prime (1)}} =  Y^{(L)}_{i,j,f^{(L)}}$. The output of the final deconvolutional layer, $\mb{Y}^{\prime (L)} \in \mathbb{R}^{m \times n \times \mathcal{F}^{(L)}}$ is then the input in a final layer meant to map this space back to the original data format $\mathbb{R}^{m \times n \times p}$. Formally the reconstructed field $\hat{\mb{X}}$ is obtained from the last convolutional step as:

\begin{equation}\label{eqn:predictions}
    \hat{X}_{i,j,p} = \tilde{g}\left(\sum_{\tilde{f}=1}^{\mathcal{F}^{\prime(L)}}\sum_{a, b=1}^{k} Y^{\prime (L)}_{i+a,j+b;\tilde{f}} \cdot w^{''}_{a,b;\tilde{f},p} + w^{''}_{0;p} \right), 
\end{equation}

\noindent where $g$ is the activation function. The final deconvolutional layer is represented as the final blue layer in Figure \ref{fig:AEArchitecture}. For the encoder and decoder in equations \eqref{eqn:encoder} and \eqref{eqn:decoder} we use the LeakyReLU activation function \citep{dubey22}, which is defined as:
\[
g(x)=\text{LeakyReLU}(x) =
\begin{cases}
    x, & \text{if } x > 0 \\
    \rho x, & \text{else}
\end{cases}
\]

\noindent where $\rho = 0.3$ for this work. This activation function is more flexible than the traditional rectified linear unit (ReLU) which assumes a value of zero for all negative inputs, for this formulation ensures there is always a non-zero gradient. For the last convolutional step in equation \eqref{eqn:predictions} we instead choose $\tilde{g}$ to be a softmax activation \citep{goo16}.

If we denote by $\mb{w}^{(\ell)}$ and $\mb{w}^{\prime (\ell)}$ the collection of all weights at layer $\ell$ for the encoder and decoder (including the bias terms), respectively, and by $\mb{w}^{''}$ the weights for the last layer, the parameters for the CAE are $\bs{\theta}_{\text{CAE}} = \{\{\mb{w}^{(\ell)}, \mb{w}^{\prime (\ell)}; \ell=1, \ldots, L\}, \mb{w}^{''}\}$.

\subsection{Time: Echo State Networks}\label{sec:forecasting}

To model the temporal structure of the encoded features $\mb{Y}$ of the CAE , we assume that the function $h_s$ in \eqref{eqn:time} is also a neural network, specifically a ESN \citep{jae01, jae07}. An ESN predicates that the weights linking the input to the hidden states and those connecting the hidden states to each other are drawn from a highly sparse spike-and-slab distribution and fixed throughout training. This stochastic approach mitigates the instability issues in gradient-based optimization methods (i.e., backpropagation) with neural networks in time. Here, we employ an ensemble of ESNs \citep{mcd17}, which involves sampling these network weights multiple times to generate a collection of predictions. We now consider the data to be dependent in time and omit the $(L)$ notation since we are modeling the output of final layer of the encoder, $\mb{Y} = \mb{Y}^{(L)}$, from the CAE. The ESN model under this framework can be described as follows \citep{mcd19b}:

\begin{subequations}\label{eqn:DESN}
\begin{flalign}
    \text{output:} & \qquad \mb{Y}_{t} = \mb{B}_{D}\mb{h}_{t, D} + \sum_{d=1}^{D-1} \mb{B}_{d}k\left(\Tilde{\mb{h}}_{t,d}\right) + \bs{\epsilon}_{t} \label{eqn:DESN1}, \\
    \text{hidden state $d$:} & \qquad \mb{h}_{t,d} = (1-\alpha)\mb{h}_{t-1,d} + \alpha\bs{\omega}_{t,d} \label{eqn:DESN2},\\
    & \qquad \bs{\omega}_{t,d} = f_{h}\left(\frac{\zeta_{d}}{|\lambda_{W_{d}}|} \mb{W}_{d} \mb{h}_{t-1,d} + \mb{W}_{d}^{\text{in}}\Tilde{\mb{h}}_{t,d-1}\right), \text{ for } d > 1 \label{eqn:DESN3},\\
    \text{reduction $d-1$:} & \qquad \Tilde{\mb{h}}_{t,d-1} \equiv Q\left(\mb{h}_{t,d-1}\right), \text{ for } d > 1 \label{eqn:DESN4},\\
    \text{input:} & \qquad \bs{\omega}_{t,1} = f_{h}\left(\frac{\zeta_{1}}{|\lambda_{W_{1}}|} \mb{W}_{1} \mb{h}_{t-1, 1} + \mb{W}_{1}^{\text{in}}\mb{z}_{t}\right) \label{eqn:DESN5},\\
    \text{matrix distribution:} & \qquad W_{d_{i,j}} = \gamma^{W_{d}}_{i,j} p(\eta_{W_{d}}) + (1-\gamma^{W_{d}}_{i,j})\delta_{0} \label{eqn:DESN6},\\
    & \qquad W^{\text{in}}_{d_{i,j}} = \gamma^{W_{d}^{\text{in}}}_{i,j} p(\eta_{W_{d}^{\text{in}}}) + (1-\gamma^{W_{d}^{\text{in}}}_{i,j})\delta_{0} \label{eqn:DESN7},\\
    & \qquad \gamma^{W_{d}}_{i,j} \text{$\thicksim$} Bern(\pi_{W_{d}}), \quad \gamma^{W_{d}^{\text{in}}}_{i,j} \text{$\thicksim$} Bern(\pi_{W_{d}^{\text{in}}}). \nonumber
\end{flalign}
\end{subequations}

\noindent In this context, $d = {1,\dots,D}$ indicates the depth or the total number of layers within the network. The output $\mb{Y}_{t}$, defined in \eqref{eqn:DESN1}, is a linear combination of terms $\mb{B}_{D}\mb{h}_{t, D}$ and $\sum_{d=1}^{D-1} \mb{B}_{d}k\left(\Tilde{\mb{h}}_{t,d}\right)$, along with an error term $\bs{\epsilon}_{t}$ which is independent and identically distributed in time as a mean zero multivariate normal distribution. Here, $\mb{h}_{t,D}$ represents a state vector of dimension $n_{h,D}$, $\Tilde{\mb{h}}_{t,d}$ is a state vector of dimension $n_{\Tilde{h},d}$, and $\mb{B}_{d}$ are matrices with parameter entries that need to be estimated. The state vector $\mb{h}_{t,d}$ is computed as a convex combination of its previous state $\mb{h}_{t-1,d}$ and a memory component $\bs{\omega}_{t,d}$, governed by the hyperparameter $\alpha$, often referred to as the \textit{leaking-rate}, as shown in equation \eqref{eqn:DESN2}.

The hidden state $\mb{h}_{t,d}$ undergoes dimensionality reduction via the function $Q(\cdot)$ in equation \eqref{eqn:DESN4}, resulting in the reduced state $\Tilde{\mb{h}}_{t,d}$, which has dimension $n_{\Tilde{h}, d}$. This dimension reduction function $Q(\cdot)$ is typically implemented using an EOF approach \citep{mcd19b, bonas23, bonas2024b}. The scaling function $k(\cdot)$ in equation \eqref{eqn:DESN1} is introduced to normalize the values of $\Tilde{\mb{h}}_{t,d}$ to be on a comparable scale to those in $\mb{h}_{t,D}$. The term $\bs{\omega}_{t,d}$ in equations \eqref{eqn:DESN3} and \eqref{eqn:DESN5} is derived using a nonlinear activation function $f_{h}$, which in this work is the hyperbolic tangent \citep{goo16}. This activation function combines the prior hidden state $\mb{h}_{t-1,d}$ with layer-specific input data. The input data, represented as $\mb{z}_{t}$, is an $n_{z}$-dimensional vector comprising past or lagged values of $\mb{Y}_{t}$, i.e., $\mb{z}_{t}=\left(\mb{Y}_{t-1},\ldots,\mb{Y}_{t-q}\right)^\top$, where $q \ge 1$ is the total number of retained lags. Alternatively, the input can be the dimensionally reduced hidden state from the previous layer, $\Tilde{\mb{h}}_{t,d-1}$.

We consider a spike-and-slab prior to inform the entries in the weight matrices $\mb{W}_{d}$ and $\mb{W}^{\text{in}}_{d}$, as shown in \eqref{eqn:DESN6} and \eqref{eqn:DESN7}. Individual matrix entries are zero with a probability of $\pi_{W_{d}}$ and $\pi_{W_{d}^{\text{in}}}$ for $\mb{W}_{d}$ and $\mb{W}^{\text{in}}_{d}$, respectively. Non-zero entries are sampled from a symmetric distribution centered around zero \citep{luk12}; in this work, $p(\cdot)$ is modeled as $\mathcal{N}(0,1)$,but other distribution choices can be chosen (see \cite{mcd17, huang2021, bonas23, bonas2024a, yoo23}).

Finally, the deep ESN must satisfy the \textit{echo state property}, which ensures that after a sufficiently long sequence, the model’s output no longer depends on its initial conditions \citep{luk12,jae07}. This property is maintained when the spectral radius (the largest eigenvalue) of $\mb{W}_{d}$, denoted as $\lambda_{W_{d}}$, remains below one. The scaling parameter $\zeta_{d} \in (0,1]$ is introduced in equations \eqref{eqn:DESN3} and \eqref{eqn:DESN5} to ensure compliance with this condition. Additionally, based on the findings of \cite{mcd19b}, we fix the hyperparameters $\pi_{W_{d}}$ and $\pi_{W_{d}^{\text{in}}}$ at 0.1, as this choice has proven robust and the overall sensitivity of the model to the choice of sparsity is negligible. In this analysis, we assume $n_{h,d}=n_h$ and $n_{\Tilde{h}, d}=n_{\Tilde{h}}$ meaning the number of nodes is fixed to be the same for every layer in the network. The model hyperparameters are collectively represented as $\bs{\theta}_{\text{ESN}}=\{n_{h}, n_{\Tilde{h}}, \zeta_{d}, \alpha, \mb{B}_d\}$, for some choice of layers $D$ where $d = 1, \dots, D$. 

\subsection{Inference}\label{sec:Inference}

Let us denote the CESAR parameters as $\bs{\theta} = \{\bs{\theta}_{\text{CAE}}, \bs{\theta}_{\text{ESN}}\}$. Inference is performed by first estimating the spatial structure with $\bs{\theta}_{\text{CAE}}$, and then conditionally on it the temporal dynamics in $\bs{\theta}_{\text{ESN}}$. 

Inference on CAE is performed by assuming no temporal dependence, so that each time point is treated as independent. We minimize the reconstruction loss, which measures the difference between the input, $\mb{X}_{t}$, and the reconstructed data, $\hat{\mb{X}}_{t}$, at the end of the decoding step. We choose the mean squared error (MSE) as a target function: $\mathcal{L}=\frac{1}{T}\sum_{t =1}^{T}\left\|\mb{X}_{i} - \hat{\mb{X}}_{i}\right\|^{2}$, even though other choices are possible. 

\noindent The MSE is minimized via gradient descent, the derivatives with respect to each parameter are computed symbolically via backpropagation and are updated by an amount $\chi$ proportional to this gradient (\textit{learning rate}). Formally, given the estimate $\bs{\theta}^{i}$ at step $i$, we have:
\begin{equation}
\bs{\theta}^{(i+1)} = \bs{\theta}^{(i)} + \chi \frac{\partial \mathcal{L}}{\partial \bs{\theta}^{(i)}}, \nonumber
\end{equation}

\noindent where $i = 1,\dots, I$ and $\chi$ denotes the number of iterations (\textit{epochs}) and the learning rate, respectively. We use the Adaptive Moment Estimation (ADAM, \cite{king14}), a popular gradient descent algorithm where $\chi$ is obtained through the past iterations to improve convergence speed. Instead of using all data, we compute the gradient by using only a subset (\textit{batch}) of size $\phi$ of the data to compute the gradients before updating the parameters \citep{leon10, rong15, mas18}. For each epoch $i$, the network processes all batches of data, updating the parameters after each batch.

Once CAE is estimated, we produce a one step ahead forecast with the CESAR model, and estimate $\bs{\theta}_{\text{ESN}}$ by computing the MSE of the forecasts across the training set. Finally, once all the parameters are learned, CESAR is used to forecast spatial maps across any lead time. We implement an iterative forecasting approach where we produce forecasts for time $T+1$ and then use forecast as part of the input data to predict the subsequent time point, $T+2$, and so on, iteratively for up to time point $T+\tau$. For this work we implement CESAR in \texttt{Python~3.12.0} and \texttt{Tensorflow~2.13} and train it across 24 cores using 2 NVIDIA A40 GPUs.

\subsection{Uncertainty Quantification}\label{sec:calibration}

The model-based framework for which we present the CESAR model in the form of equation \eqref{eqn:statespace} allows the uncertainty to be naturally provided. For this work we quantify uncertainty from the proposed CESAR in space and time separately. That is, we \textit{calibrate} the forecasts from CESAR by first computing the spatial forecasting uncertainty stemming from the CAE portion of the model and then we compute the temporal uncertainty from the ESN portion of the model. We aim at providing a calibrated uncertainty so that, e.g., a 95\% prediction interval must cover the true (unobserved) value $\sim$95\% of the time. 

To quantify the spatial uncertainty, we rely on \textit{dropout} \citep{gal16}, an approach which randomly forces some parameters in the network to be zero during the model's training. Formally, we can reformulate the CAE from equations \eqref{eqn:encoder}, \eqref{eqn:decoder} and \eqref{eqn:predictions} to incorporate this random probability of parameter exclusion. For example, we rewrite equation \eqref{eqn:encoder} as:
\begin{equation}\label{eqn:dropout_encoder}
    Y^{(\ell)}_{i,j;f} = g\left(\sum_{\tilde{f}=1}^{\mathcal{F}^{(\ell-1)}} \sum_{a, b=1}^k Y^{(\ell-1)}_{\xi i+a,\xi j+b;\tilde{f}} \cdot w^{(\ell)}_{a,b;\tilde{f},f} \cdot p^{(\ell)}_{a,b;\Tilde{f},f} + w^{(\ell)}_{0;f} \cdot p^{(\ell)}_{0;f} \right).
\end{equation}

\noindent where $p^{(\ell)}_{a,b;\Tilde{f},f}, p^{(\ell)}_{0;f}\thicksim Bern(\varphi)$ and $1-\varphi$ represents the `dropout rate' or the percentage of parameters which are removed from the network. A similar formulation is provided for equations \eqref{eqn:decoder} and \eqref{eqn:predictions}. This random deletion of parameters from the network allows to generate an ensemble of predictions with different parts of the network turned off each time. From this collection of predictions we are then able to calculate quantiles to estimate the uncertainty surrounding the spatial component of the model. It will be shown throughout the simulation study and application how this approach yields proper estimation of the uncertainty for the spatial component of the forecasts.

For the temporal component we assess the uncertainty via an ensemble-based approach which leverages on the stochastic nature of the ESN. Indeed, one can sample the network weights multiple times via their spike-and-slab prior in equations \eqref{eqn:DESN6} and \eqref{eqn:DESN7}. Formally we can denote these draws of weights for equations \eqref{eqn:DESN6} and \eqref{eqn:DESN7} using the notation $W^{(\iota)}_{d_{i,j}}$ and $W^{\text{in}, (\iota)}_{d_{i,j}}$, respectively. Here, $\iota = 1, \dots, \mathcal{I}$ and $\iota$ represents the index or specific draw of weights then used to generate forecasts, $\mb{Y}^{(\iota)}_{t}$, and $\mathcal{I}$ is the total number of independent draws taken from the spike-and-slab prior. These multiple draws of weights the yield an ensemble of $\mathcal{I}$ total forecasts. From this ensemble of forecasts one can then produce uncertainty estimates similarly to the dropout approach detailed above by computing quantiles.

\section{Simulation Study}\label{sec:simstudy}

We perform a simulation study with a two dimensional Burgers' equation \citep{burger48} which we introduce in Section \ref{sec:Burger2D}, in order to assess:
\begin{enumerate}
    \item the ability of CAE against other spatial dimension reduction approaches to extract spatial features in Section \ref{sec:Burger2DResults},
    \item the forecasting skills of CESAR versus other time series methods in Section \ref{sec:forcsim},
    \item uncertainty quantification in Section \ref{sec:uqsim}.
\end{enumerate}

\subsection{Simulated data}\label{sec:Burger2D}


We consider a two dimensional Burgers' equation, which is a special case of Navier Stokes, the basic equations for fluid dynamics. The equation assumes a two dimensional fluid with viscosity $\nu>0$ \citep{burger48, bate15} and velocity $\mb{u} = (u(x,y,t), v(x,y,t))$ at some location $(x,y)$ and time $t$\footnote{It could be confusing to have the first element with the same letter of the vector, but we preferred to keep the standard notation from fluid dynamics}. Using the notation from Section \ref{sec:spacetime}, the spatio-temporal process we are analyzing would be $\mb{X}_{t} = \mb{u}(x,y,t)$. We define the Burgers' equation on a unit square with periodic boundary conditions as follows \citep{gao17, gen20}:
\begin{subequations}\label{eqn:Full2DBurger}
\begin{gather}
\mb{u}_{t} + \mb{u} \cdot \nabla\mb{u} = \nu \Delta \mb{u}, \\
\mb{u}(x, 0, t) = \mb{u}(x,1,t), \quad \mb{u}(0, y, t) = \mb{u}(1,y,t)
\end{gather}
\end{subequations}
\noindent where $\mb{u}_{t}$ is the time derivative, $(x, y) \in [0,1]$ and $t \in [0,1]$. We rely on data from \cite{gen20}, where the fluid viscosity is fixed to $\nu = 0.005$, and the domain is a $m\times n = 64\times 64$ grid, so that the unit square is solved on a grid of size $1/64\times 1/64$. For time, we consider a discretization in steps of $0.01$ in time for $T=101$ total time steps.

\begin{figure}[!tb]
\centering
\includegraphics[width = 15cm]{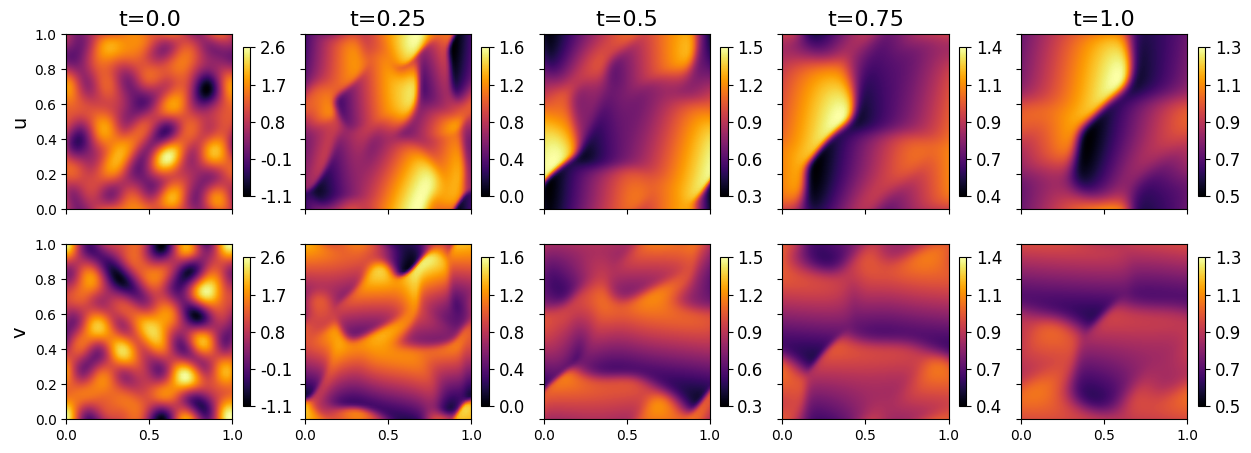}
\caption{Simulation from the two dimensional Burgers' equation \eqref{eqn:Full2DBurger} with periodic boundary conditions and initial conditions as a Fourier series \eqref{eqn:Fourier4Order}. This figure depicts the two velocity fields for $t = \{0,0.25,\dots,1\}$.}
\label{fig:BurgerSimExample}
\end{figure}

The initial condition is a Fourier series:
\begin{subequations}\label{eqn:Fourier4Order}
\begin{gather}
    \bs{\psi}(x,y) = \sum_{a = -4}^{4}\sum_{b=-4}^{4}\bs{\alpha}_{ab}\text{sin}(2\pi\left(ax+by\right))+\bs{\beta}_{ab}\text{cos}(2\pi\left(ax+by\right)),\\
    \mb{u}(x,y,0) = \frac{2\bs{\psi}(x,y)}{\text{max}_{x,y}|\bs{\psi}(x,y)|} + \bs{\eta},
\end{gather}
\end{subequations}

\noindent and we consider $n_{\text{sim}}=10$ simulations, each with a different realization from $\bs{\alpha}_{ab}, \bs{\beta}_{ab} \thicksim \mathcal{N}(\mb{0},\mb{I})$ and $\bs{\eta} \thicksim \mathcal{U}(-1,1) \in \mathbb{R}^2$. From Figure \ref{fig:BurgerSimExample} we can see an example of one simulation at selected time points. The velocity field for both components is initially excited from the initial conditions, but gradually dissipates energy due to the viscosity so that the small scale features gradually vanish.

We were able to train CESAR in less than 5 minutes per simulated dataset. This computational time could easily be further improved by increasing the number of cores or GPUs.

\subsection{Space: CAE Reconstruction}\label{sec:Burger2DResults}

We choose a CAE with $L=3$ layers in equations \eqref{eqn:encoder} and \eqref{eqn:decoder}, $k=3$ filter size and stride $\xi = 2$. For the encoder block we set the number of filters at $\mathcal{F}=\{16, 32, 64\}$, while the reversed order is used for the decoder. This configuration allows us to have a latent space of dimension comparable to all other competing methods, so that the comparison is as fair as possible. We use $I=500$ epochs and a batch size of $\phi = 2$ to train the CAE. The first 80 time points were used as training set to predict the velocity field for the last $\tau=21$ time points. 

We use PCA, Inverse Distance Weighting (IDW), Kriging and FRK (with Mat\'ern covariance) as competing methods, and we provide details about their configuration in the supplementary material. The maximum number of principal components from the PCA were retained for reconstruction in order to generate the best possible and representative mapping. We compare the spatial reconstructions in terms of the MSE ($\times 10^{-3}$) in Table \ref{tbl:SpatialMSE}. This table depicts the median performance for each method across space and the $n_{\text{sim}}=10$ simulated data with the interquartile range (IQR) reported in parenthesis. The CAE drastically outperforms the all other methods, with a median MSE of 0.27 (IQR = 0.87) for spatial reconstruction which translates to an 35.7\% improvement over the next best method, PCA, which returns a median MSE of 0.42 (1.85).

\subsection{Time: ESN Forecasting skills}\label{sec:forcsim}

The optimal architecture for the ESN was with $D=1$ layers in equations \eqref{eqn:DESN}, with $n_{h} = 64$ hidden nodes, see supplementary material for the sensitivity study with respect to these choices. The other parameters, $\zeta$, $\alpha$, and $\mb{B}$ in equation \eqref{eqn:DESN}, are learned.

As alternative methods, we consider two popular approaches to time series modeling: ARIMA and Long Short-Term Memory (LSTM, \cite{hoch97}) network. Additionally, we also use a persistence, a simple method which assumes that all future forecasts assume the value of the final observed data in the training set. From Table \ref{tbl:TemporalMSE} it seen how the ESN model produces the best MSE when coupled with the CAE versus ARIMA, persistence and an LSTM, for it returns a metric of 3.81 (11.38) which yields a 11.2\%, 14.4\%  and 71.8\% improvement over ARIMA, persistence, and the LSTM, respectively. The poor performance of the LSTM approach can at least partly attributed to the lack of a large training set available for the model.

\begin{figure}[!tb]
\centering
\includegraphics[width = 13cm]{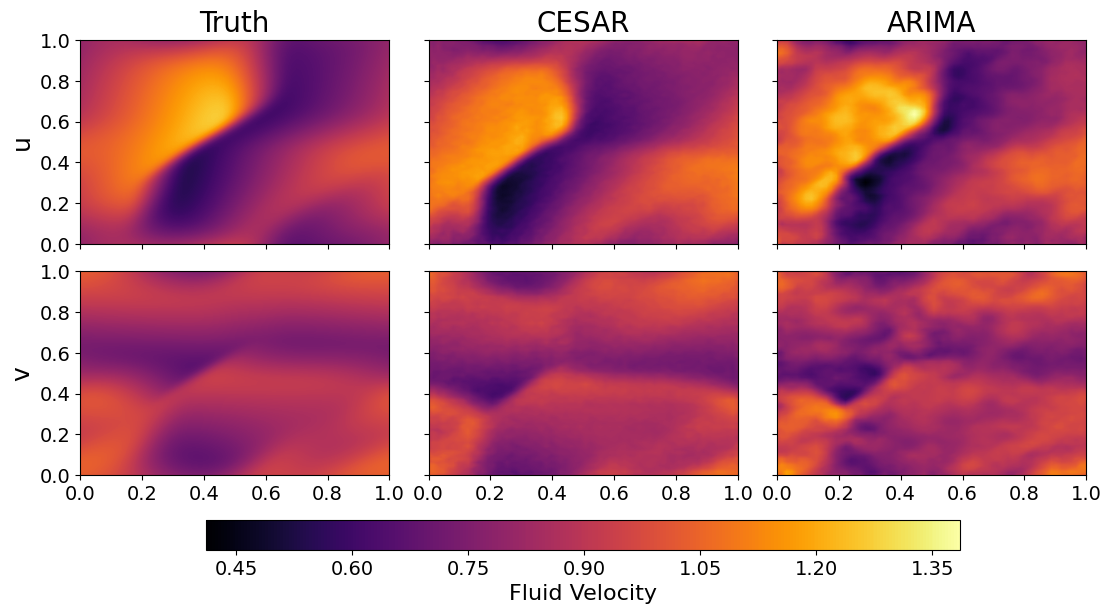}
\caption{Comparison between the true data and the CESAR and ARIMA forecasts for both velocity fields of the two dimensional Burgers' equation \eqref{eqn:Full2DBurger}. For consistency, the same simulation depicted in Figure \ref{fig:BurgerSimExample} is used. This figure shows the velocity field averaged across the final $\tau=21$ time points.}
\label{fig:BurgerForecasts}
\end{figure}

To visually assess the performance of CESAR, we show the average forecasts for a single simulation in Figure \ref{fig:BurgerForecasts}. It can be seen how CESAR yields a smooth and visually accurate representation of the simulated data as opposed to ARIMA, which instead provides a more noisy forecast with more local artifacts. This figure and Table \ref{tbl:SpatialMSE} support the use of the proposed CESAR approach which combines a CAE as a method for nonlinear feature extraction and dimension reduction coupled with an ESN for time series forecasting.

\subsection{Uncertainty Quantification}\label{sec:uqsim}

In order to assess the uncertainty of the spatial reconstruction, we rely on \textit{dropout} as detailed in Section \ref{sec:calibration}. The percentage of parameters dropped was chosen such that the prediction intervals was as calibrated as possible. The optimal choice for the dropout rate was 0.79 or $\varphi = 0.21$ in Section \ref{sec:calibration}. The sensitivity study with respect to the dropout rate is deferred to the supplementary material. For the temporal component, we implement the ensemble-based approach detailed in Section \ref{sec:calibration}. 

Table \ref{tbl:AllUQ} show the empirical coverage on the last $\tau=21$ points for a nominal 95\%, 90\% and 80\% confidence interval for both the spatial and temporal component, where the IQR across the $n_{\text{sim}}=10$ simulated datasets is shown in parenthesis. For the temporal component, the coverage is evaluated marginally across each spatial location (column 3) and jointly using the grand mean of the data (column 4). From this table it is readily apparent how the uncertainty is properly estimated both spatially and temporally. Specifically, the average coverage discrepancy across the three intervals is approximately 3\% for each of the methods.

\begin{table}[!tb]
\centering
\begin{tabular}{ccc}
\toprule
Method & 2D Burgers' ($\times 10^{-3}$) & WRF ($\times 10^{-1}$)\\
\midrule
CAE & 0.27 (0.87) & 0.17 (0.18) \\
PCA & 0.42 (1.85) & 7.00 (4.02)\\
IDW & 3.41 (6.26) & 9.93 (6.63) \\
Kriging & 0.48 (1.73) & 8.80 (5.75) \\
FRK & 1.14 (3.06) & 9.47 (5.99) \\
\bottomrule
\end{tabular}
\caption{Median (across space) forecasting MSE for each method for both the simulated two dimensional (2D) Burgers' equation and WRF wind speed ($ms^{-1}$) data. The IQR is reported in parenthesis.}
\label{tbl:SpatialMSE}
\end{table}

\begin{table}[!tb]
\centering
\begin{tabular}{ccc}
\toprule
Method & 2D Burgers' ($\times 10^{-3}$) & WRF ($\times 10^{-1}$)\\
\midrule
ARIMA & 4.29 (13.20) & 29.09 (23.21) \\
LSTM & 13.53 (42.69) & 23.12 (11.46)\\
CESAR & 3.81 (11.38) &  19.21 (8.60) \\
Persistence & 4.45 (11.69) & 40.53 (36.71) \\
\bottomrule
\end{tabular}
\caption{Median iterative 1-step ahead forecasting MSE across space for each method for both the simulated two dimensional (2D) Burgers' equation and WRF wind speed ($ms^{-1}$) data. Each time series method was used in conjunction with the CAE. The IQR is reported in parenthesis.}
\label{tbl:TemporalMSE}
\end{table}

\begin{table}[!tb]
\centering
\begin{tabular}{ccccccc}
\toprule
  \multirow{4}{*}{Coverage} & \multicolumn{3}{c}{2D Burgers'} & \multicolumn{3}{c}{WRF} \\
  \cmidrule(lr){2-7}
        & \multirow{2}{*}{Spatial} & \multicolumn{2}{c}{Temporal} & \multirow{2}{*}{Spatial} & \multicolumn{2}{c}{Temporal} \\
\cmidrule(lr){3-4}\cmidrule(lr){6-7}
       &  & Marginal & Grand Mean & & Marginal & Grand Mean\\
\midrule
    95\% & 93.9 (9.1) & 92.9 (3.9) & 99.0 (2.9) & 92.6 (6.9) & 95.7 (4.9) & 95.8 \\
    90\% & 91.4 (10.9) & 86.9 (7.1) & 91.4 (11.6) & 89.4 (8.4) & 91.0 (7.8) & 87.5 \\
    80\% & 86.4 (13.6) & 75.4 (11.7) & 84.0 (18.8) & 83.6 (10.7) & 80.4 (11.0) & 75.0  \\
\bottomrule
\end{tabular}
\caption{Empirical coverage of confidence intervals from three nominal levels. The spatial coverage uses dropout and is shown in columns 2 and 5 for the 2D Burgers' equation and the WRF simulation, respectively. The temporal coverage using the ensemble-based method is shown in the remaining columns for both the 2D Burgers' equation data and WRF wind speed output. For the 2D Burgers' equation data an average across simulated datasets is shown with the standard deviation reported in parenthesis. For the WRF wind speed data an average across spatial locations is shown with the standard deviation reported in parenthesis}
\label{tbl:AllUQ}
\end{table}

\section{Application}\label{sec:appl}

We now apply the proposed CESAR approach to forecasting wind speed and energy from the WRF simulation as detailed in Section \ref{sec:data}. We split the data by using the first $T=217$ hours (9 days) for training and the last $\tau=24$ hours for testing. Inference for CESAR was achieved in less than 30 minutes, and the increase in computational time versus the simulation study ($\approx$ 5 minutes) can be attributed to the increased number of time points compared to the simulation study.

\subsection{Wind Speed Prediction and Uncertainty Quantification}\label{sec:WRFResults}

We choose to implement the CESAR model with an encoder and decoder blocks each comprising  $L=3$ convolutional layers, each with filters of size $k=3$ with a stride of $\xi=2$, thus reducing the input dimensions by half at each layer. The encoder block's layers were designed with filter counts of $\mathcal{F}=\{32, 64, 128\}$ respectively, and the same reversed number with for the decoder. The ESN component used $D=1$ layers, similar to that of the simulation study, and $n_{h} = 128$ hidden nodes. We use $I = 1000$ epochs with a batch size of $\phi = 10$.

\begin{figure}[!tb]
\centering
\includegraphics[width = 13cm]{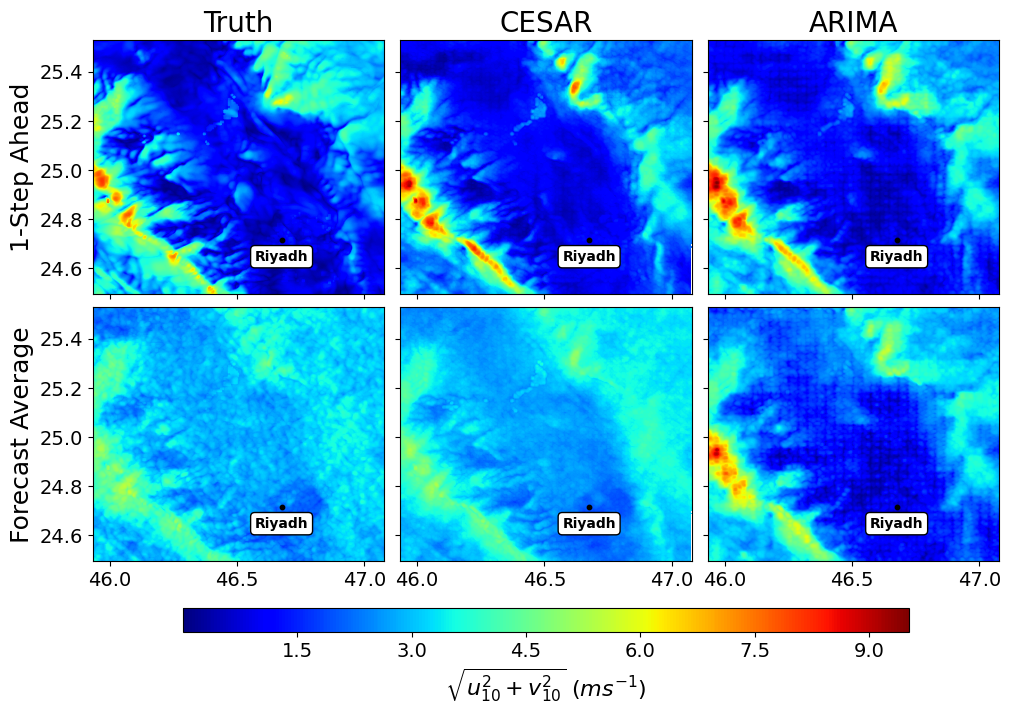}
\caption{Forecast of the WRF wind speed data for proposed CESAR method and the ARIMA approaches. This figure depicts the 1-step ahead forecasts in the first row and the average wind speed forecasts across space for $\tau=24$ hours in the second row.}
\label{fig:WSSpatial}
\end{figure}

Table \ref{tbl:SpatialMSE} shows a comparison between the spatial methods in terms median spatial MSE across time, with the IQR is reported in parenthesis. Similarly to the simulation study, the CAE significantly outperforms all of the other approaches. Specifically, the CAE achieves a median MSE ($\times 10^{-1}$ $ms^{-1}$) of 0.17 (0.18) for spatial reconstruction, a 97.6\% improvement over the next best method, the PCA, which yielded a MSE of 7.00 (4.02). Table \ref{tbl:TemporalMSE} shows instead how the temporal part of CESAR (i.e., the ESN) produces the best MSE when compared with CAR with ARIMA, persistence, and LSTM. Specifically, CESAR returns a MSE ($\times 10^{-1}$ $ms^{-1}$) of 19.21 (8.60) which then translates to a 34.0\%, 52.6\%  and 16.9\% improvement over ARIMA, persistence and the LSTM, respectively. The improved performance of CESAR against the results in Section \ref{sec:simstudy} can be at least partly attributed to the increased amount of training data available for the model. We further visually illustrate the improved forecasting skills with Figure \ref{fig:WSSpatial}, which shows both the 1-step ahead (row 1) and average forecasts in time (row 2) across the training set. CESAR produces a more accurate representation of the true field in comparison to ARIMA. While for 1-step ahead forecast is relatively similar, the average forecast clearly shows overestimation of ARIMA, especially in the plateau of Jabal Tuwaiq, the escarpment west of Riyadh. For the Najd plateau where the city resides, the wind is instead underestimated. 

Given the temporal behavior of the boundary layer, especially in a desertic region as the one surrounding Riyadh, the forecasting skills are expected to show strong differences depending on day and night. In the supplementary material we break down CESAR's forecasting skills for daytime and nighttime hours. As mentioned in Section \ref{sec:data}, the atmosphere is stable during the nighttime hours whereas it exhibits strong mixing and turbulent behavior during the daytime hours. This causes a nightly stable and spatially structured signal (Figure \ref{fig:WSDataPlot}A) as opposed to a daily complex and almost chaotic patterns (Figure \ref{fig:WSDataPlot}B). As discussed in the supplementary Section 5 and Figure S1, the noisy daytime structure results in suboptimal forecasts for the CESAR model (while still outperforming the next best model presented in Table \ref{tbl:TemporalMSE}, the LSTM), whereas during the night the CESAR forecast is able to capture the underlying spatial structure. This is particularly important for wind energy applications, specifically wind farm planning/design and wind power forecasting, as the wind speed tend to peak at nighttime in the region. 

Lastly, as in the simulation study in Section \ref{sec:simstudy}, we choose to estimate the uncertainty surrounding the spatial reconstruction using dropout which optimal rate of $\varphi = 0.3$ and we assess the forecast uncertainty using the ensemble forecasting based approach, both of which are detailed in Section \ref{sec:calibration}. Table \ref{tbl:AllUQ} depicts the empirical coverage of the 95\%, 90\% and 80\% levels for each of these approaches, where a average across space is shown with the standard deviation in parenthesis. We again calculate the marginal temporal uncertainty across each spatial location (column 6) and joint temporal uncertainty using the grand mean of the data and forecasts (column 7). This table clearly shows how the uncertainty is properly estimated both spatially and temporally: the average coverage discrepancy across the three intervals is smaller than that reported for the simulation study, approximately 2\% for each of the methods.

\subsection{Wind Power Estimation}\label{sec:WRFPower}

In this section we describe how we can use the forecasts generated by CESAR to extrapolate the wind speed at a higher altitude (i.e., the turbine hub height) and then estimate the wind power generation using wind turbines at each spatial location. In Section \ref{sec:WSExtrap} we detail the approach to wind speed extrapolation and in Section \ref{sec:WindPower} we detail the turbine specifications and the wind power estimation for the spatial domain.

\subsubsection{Wind Speed Extrapolation}\label{sec:WSExtrap}

In this section we denote the wind speed across the domain at some time point $t$ and height $h$, as $\mb{X}_{h,t}$. Wind turbines are typically within the height range of 80m to 110m, so we need to extrapolate the 10m forecasts from CESAR to this range. Wind speed extrapolation at hub height $\Tilde{h}$ is a common approach for assessing wind energy \citep{gual19, zhang24}, and for this work, we choose the most popular model, the \textit{power law}, which assumes that:
\begin{equation}
    \mb{X}_{\Tilde{h},t} = \mb{X}_{h,t}\left(\frac{\Tilde{h}}{h}\right)^{\kappa}.
\end{equation}

\noindent Here, $\kappa$ denotes the \textit{shear coefficient} which controls the magnitude of the change in average wind speed with respect the change in height \citep{crippa21}. As is the standard in the power law literature, we assume the shear coefficient $\kappa = \frac{1}{7}$ constant in space and time, a value corresponding to a relatively flat surface and neutral atmospheric conditions \citep{peter78}. A few recent studies have investigated more robust general methods to estimate the shear coefficient for multi-year runs (see \cite{zhang24, crippa21} for examples). For this work, however, we have a short time frame of 10 days at hourly scale and a small spatial region so, so we opted to use the standard model with constant $\kappa$ in space and time. Once the wind speed is extrapolated to hub height, we convert the wind speed estimates to power estimates using \textit{power curves} of the desired turbine. 

\subsubsection{Wind Power Generation}\label{sec:WindPower}

The estimated wind speed at the turbine hub height should then be converted to wind power estimates (in kW) using \textit{power curves}. A power curve describes the power generated by a turbine given a specific wind speed, and is zero until a minimum speed makes the blade turbines rotate (cut-in). The power from the turbines increases as the wind speed increases until it reaches a maximum power generation (cut-out) at which point it remains constant. For this work we a turbine make and model that \cite{giani20} identified being optimal for the region of interest: the Nordex N100-2500, which has a maximum power generation of 2500kW and has a hub height of 80m. Using the power curve and height of this turbine we can compute the estimated power for each grid cell in the spatial domain, assuming a single turbine is present in each cell. For this section we only consider the one step aheat forecasts at 2016/07/31 01:00:00 UTC, for in the context of wind power estimation these forecasts are important for electrical grid management and turbine operations \citep{gieb11}.

\begin{figure}[!tb]
\centering
\includegraphics[width = 12.5cm]{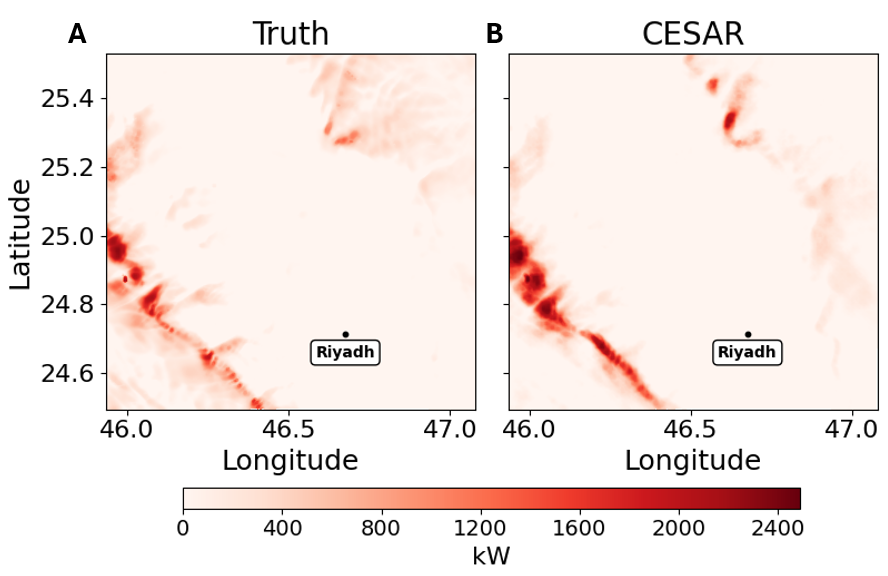}
\caption{Wind power forecasts from CESAR for 2016/07/31 01:00:00 UTC for the domain of interest assuming a single Nordex N100-2500 turbine is present in each grid cell at a hub height of 80m.}
\label{fig:WRFPower}
\end{figure}

Figure \ref{fig:WRFPower} depicts the true wind power and the forecasted one from CESAR for each spatial location. CESAR largely produces accurate estimates of wind power for the entire spatial domain and shows high wind power estimates west of the city of Riyadh. In fact, the average forecasted power across the entire spatial domain (with standard deviation) is 78.3kW (269.3kW) versus the true wind power of 83.2kW (213.8kW).

\section{Discussion}\label{sec:discuss}

\quad In this work, we have introduced a new approach to wind forecasting over a large spatial domain by combining spatial modeling with a convolutional autoencoder and temporal modeling with an echo state network and formulated it as a hierarchical space-time model. Evidence from both the simulation study and the application indicates that our proposed CESAR model effectively extracts the spatial features of the high-resolution wind speed data, captures complex nonlinear dynamics and produces reliable forecast estimates. Additionally, CESAR is computationally efficient and enables uncertainty quantification through an ensemble-based approach. While we applied this approach to wind speed over a region around Riyadh (Saudi Arabia), this methodology is broadly applicable to any spatio-temporal data involving complex, nonlinear patterns and relatively limited samples in time.

The main methodological innovation of this work is in merging two powerful frameworks in deep learning spatial and temporal data, and cast the resulting method as a hierarchical space-time model. First, we employ a CAE to reduce the dimensionality and extract latent features from high-dimensional wind fields. Second, we integrate this with an ESN, an approach capable of learning complex temporal dependencies even with limited training data, to produce accurate wind speed forecasts. CESAR is evaluated on both a 2D Burgers' equation and high-resolution weather data, with the results demonstrating the superiority of the proposed approach yielding improvements in terms of the spatial MSE up to 97.6\% and temporal MSE up to 71.8\% over the other competing methods.

In its current formulation, CESAR is limited to data on a regular grid, a feature which is very common on simulated data, but not with observational data. If the sampling is irregular, more general CAEs relying on graph convolutional networks \citep{wan24} would be necessary. Also, CAE do not assume an underlying continuous spatial process, so spatial interpolation cannot be performed, at least with the currently defined model. While this is not an issue for simulated data on a regular grid for which forecast is sought, some environmental applications may require interpolation (\textit{downscaling}) and hence interest in the conditional distribution at unsampled locations. Finally, this work aims at providing operational forecasts for wind energy to allow management of the energy grid in Saudi Arabia, especially in the Riyadh Province. Since, at the time of writing, the country does not have an automatic, continuously updated data product for weather forecasting such as the High-Resolution Rapid Refresh  \citep{hrrr22} in the United States, we relied on \textit{ad hoc} WRF simulations, with the long-term goal of integrating our modeling work in the country's emerging operational forecasting cyberinfrastructure.

\section{Acknowledgments}\label{sec:acknowledgments}

This study is based upon work supported by the National Science Foundation under Grant No. 2236504 to PC and No. 2347239 to SC and MB, and by the King Abdullah University of Science and Technology (KAUST) Office of Sponsored Research (OSR) under Award ORFS-2022-CRG11-5069.2. 

\begin{center}
\textbf{\large Supplementary Material}
\end{center}

\noindent The \texttt{Python} code to apply the CESAR approach in Sections \ref{sec:methods} on a simulated 2D Burgers' equation, along with the data, can be found in the following GitHub repository: \url{github.com/MBonasND/2025CESAR}.

\bibliographystyle{chicago}
\bibliography{references}

\end{document}